\documentclass[twocolumn]{el-author}
\usepackage[dvipsnames]{xcolor}
\usepackage{amsmath}

\newcommand{\ua}{\uparrow}
\newcommand{\nc}{\newcommand}
\nc{\da}{\downarrow} \nc{\hc}{\hat{c}} \nc{\hS}{\hat{S}}
\nc{\bra}{\langle} \nc{\ket}{\rangle} \nc{\eq}{equation (\ref}
\nc{\h}{\hat} \nc{\hT}{\h{T}}\nc{\be}{\begin{eqnarray}}
\nc{\ee}{\end{eqnarray}}\nc{\rd}{\textrm{d}}\nc{\e}{eqnarray}\nc{\hR}{\hat{R}}\nc{\Tr}{\mathrm{Tr}}
\nc{\tS}{\tilde{S}}\nc{\tr}{\mathrm{tr}}\nc{\8}{\infty}\nc{\lgs}{\bra\ua,\phi|}\nc{\rgs}{|\ua,\phi\ket}
\nc{\hU}{\hat{U}}\nc{\lfs}{\bra\phi|}\nc{\rfs}{|\phi\ket}\nc{\hZ}{\hat{Z}}\nc{\hd}{\hat{d}}\nc{\mD}{\mathcal{D}}
\nc{\bd}{\bar{d}}\nc{\bc}{\bar{c}}\nc{\mc}{\mathcal}\nc{\ea}{eqnarray}\nc{\mG}{\mathcal{G}}\nc{\bce}{\begin{center}}
\nc{\ece}{\end{center}}
\date{26th May 2022}

\usepackage{graphicx}

\begin{document}

\title{An Integrated, Phase-Controlled Power Divider for Metasurface Array Antennas}

\author{D. Banerjee, A.V. Diebold, D.R. Smith, and M. Boyarsky}

\abstract{We present the design of a phased, modulated, power distribution network for metasurface array antennas. The specific metasurface array design comprises a series of waveguides, each feeding a sub-array of dynamically tunable metamaterial elements that radiate at microwave (X-band) frequencies. To remain a one-port device, the composite array requires a power divider that can excite each branch waveguide with equal power. Further, the power divider must also apply a prescribed phase shift to each branch waveguide to mitigate metasurface-specific grating lobes. The presented design features hollow metal waveguides, with a conducting sheet patterned with metamaterial irises serving as the upper wall. A single hollow metal main waveguide feed is used to couple power to each of the branch waveguide. Carefully oriented slots control both the magnitude and phase of the field distributed from the main feed to each branch. All waveguides can be machined from a single block of metal, with a metamaterial-patterned printed circuit board laminated to the top of the assembly. The resulting design is mechanically robust and capable of high power operation, while retaining the low-cost, low-power consumption, lightweight and low-profile features common in metasurface antenna designs.}

\maketitle

\section{Introduction}

The advent of modern microwave devices for wireless communication, remote sensing, and security screening has continued the widespread need for microwave antennas. Several antenna architectures exist, including phased arrays and reflector dish systems, but the cost versus performance of these solutions restrict their large-scale adoption, motivating further development of electronically-steered systems. Metasurface array antennas have received increasing attention as beam-steering antennas that can be fabricated using printed circuit board (PCB) and similarly mature manufacturing approaches [1]-[5]. These antennas can deliver high-quality beam forming from a low cost, lightweight, and planar platform, making them an attractive option for many microwave applications.

\par As a consequence of their inherently passive nature (i.e., no externally powered elements such as phase shifters or amplifiers in the tuning network), metasurface antennas come with a particular set of design challenges. A common metasurface array antenna architecture \cite{Mike_1} comprises a series of waveguide-fed metasurface antennas, each with a set of resonant, metamaterial elements etched into the upper conductor of the waveguide structure. By changing a control stimulus, the elements can be tuned, either continuously or between two states, providing a means of beam forming over the two angular directions. Metamaterial element control, however, is limited in that a single, passive, resonant element can allow at most $180^\circ$ of phase tuning and the phase and magnitude response are linked. These limitations can lead to metasurface-specific grating lobes. However, introducing variation among the phases of the feed waves exciting each waveguide can mitigate these grating lobes. This prescribed phase variation at the feed points represents another requirement in the design of a metasurface antenna system \cite{Mike_1,Mike_2}.


\par While the literature is replete with diverse phased array antenna topologies, less work has been reported on the design of the matching feed network. A methodology to feed phased array antennas was proposed by \cite{Bocio}, which incorporated coherent beam-forming networks to provide phase diversity. The network was frequency and power limited, as it was implemented for Yagi arrays in the S-band. Other existing alternatives for power dividers include slotted waveguide planar structures \cite{Wang}, which exhibit form-factor limitations and require a separate antenna housing. Substrate Integrated Waveguide divider designs \cite{Song, Eom} also work well, but are limited by the maximum power handling capacity. Additionally, cavity resonator-based dividers have been demonstrated \cite{Mohammed}, but are heavy and bulky. With most microwave applications requiring phase-constant power division, metasurface antennas present a unique design requirement of a phase-controlled power divider.

\par
\begin{figure}[!t]
\centering{\includegraphics[width=63mm]{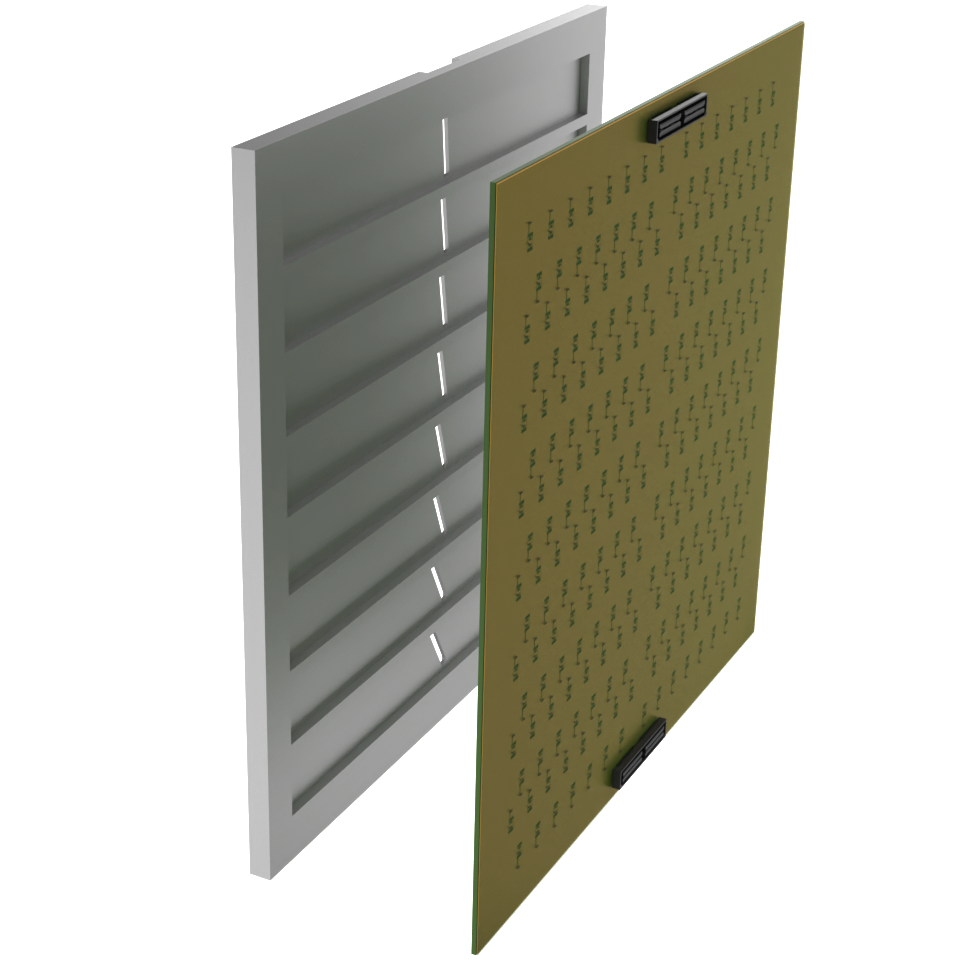}} \vspace{-0.1in}
\caption{Metasurface antenna with an integrated, phase-controlled power distribution network \vspace{-0.1in}
\label{placement}}
\end{figure}

 Here, we consider the design of a power distribution network for metasurface antennas that simultaneously provides equal power division and prescribed phase control, while also allowing higher power operation. The design performance, illustrated in Fig. \ref{placement}, is validated through full-wave simulations. The frequency of operation is chosen to be $f_0=9.75\mbox{GHz}$ with a target bandwidth extending from $9.2\mbox{GHz}$ through $10.2\mbox{GHz}$. This paper describes of the design structure, details of parameter optimization, and simulation results. While we present our design in the context of one particular metasurface antenna architecture, the design principle can be generalized to many varieties of metasurface array antennas or other applications requiring power division and phase control simultaneously. \vspace{-0.1in}

\section{Structure of the Proposed Power Distribution Network}


\begin{figure}[!ht]
\centering{\includegraphics[width=85mm]{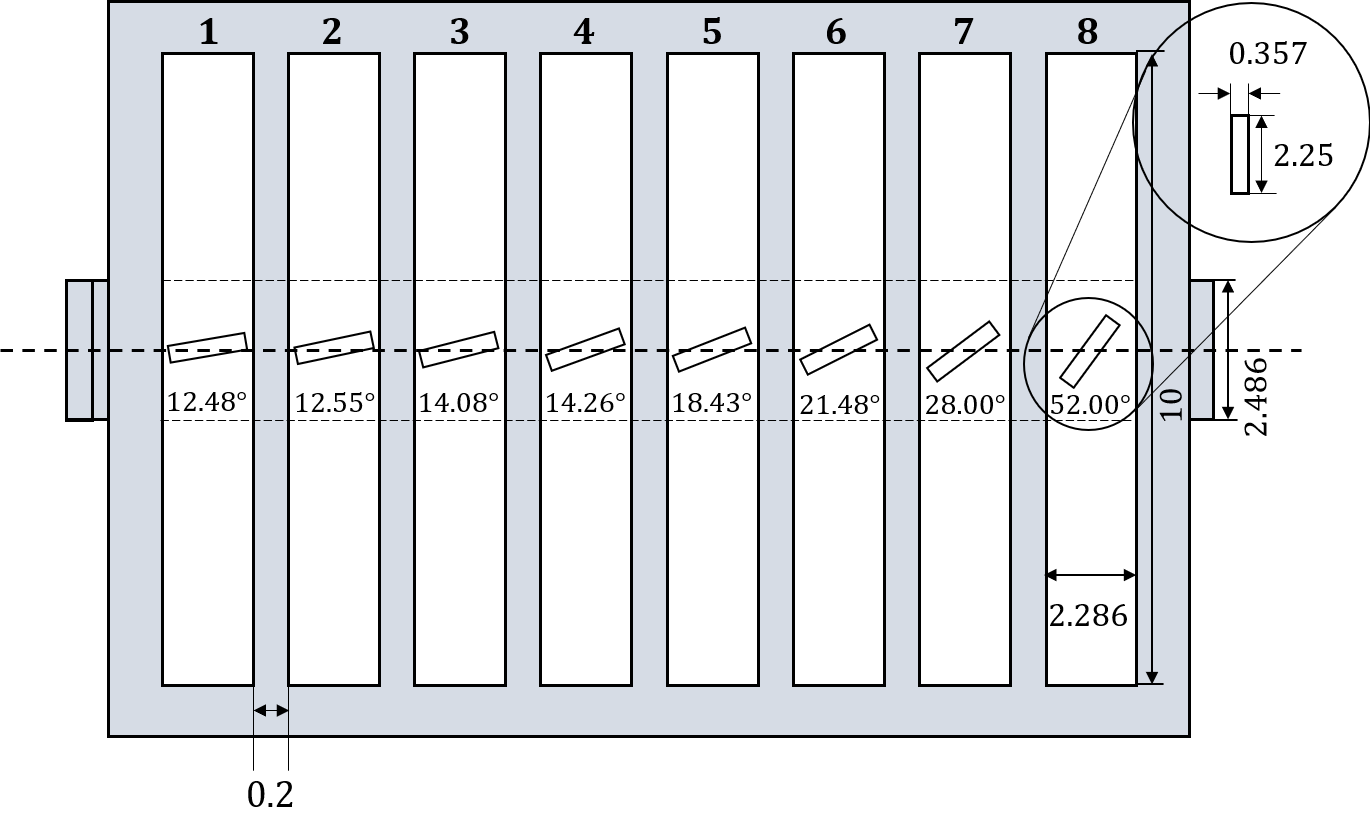}} \vspace{-0.1in}
\caption{Structure of the proposed phased power distribution network with slot orientation and dimensions. All measurement units are in cm
\label{CST_structure}}
\end{figure}

The metasurface array aperture is targeted to have an area of $25$ x $25$ $cm^2$, with the ability to operate at high power levels in the X band. To accommodate high RF power levels, hollow metal waveguides are used, with a thin PCB laminated to the top of the metal structure. The metamaterial elements and control components are assumed to be contained within the thin PCB. To fill the antenna area, eight waveguides are arrayed adjacently. Each waveguide requires a unique relative phase to avoid grating lobes. The proposed structure is illustrated in Fig. \ref{CST_structure}.

\par Power is coupled from the main feed line to the branch waveguides through coupling slots \cite{Rengarajan} to ensure an equal-power split. The waveguides can be either terminated or shorted at the end. The waveguides are the same width as a WR-90 waveguide ($22.86$ mm), but have a non-standard waveguide height of $0.3\mbox{cm}$ to reduce the structure thickness and overall antenna profile. The waveguides are spaced apart by $0.2\mbox{cm}$.

\par
\begin{figure}[!ht]
\centering{\includegraphics[width=70mm]{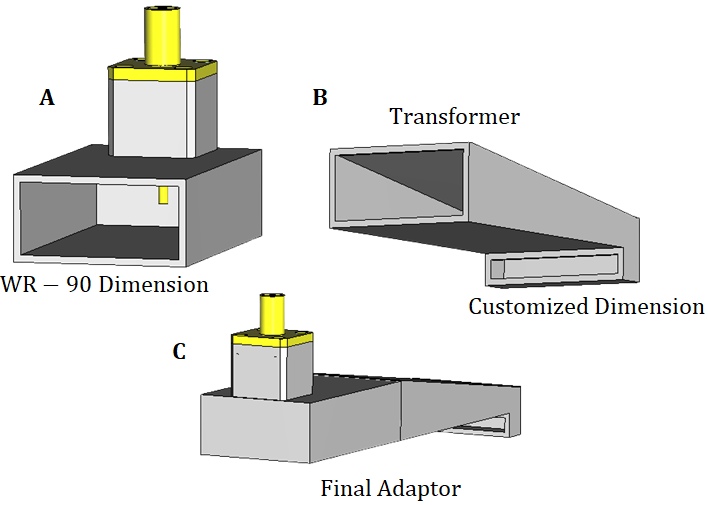}} \vspace{-0.1in}
\caption{Adapter transformation from standard $50\Omega$ SMA to customized waveguide dimension input \vspace{-0.1in}
\label{adapter}}
\end{figure}

 The non-standard waveguide height introduces the need for an adapter that can transition from a standard WR-90 input. here we assume the input is a standard SMA connector, which must be transitioned to the custom waveguide dimensions, as illustrated in Fig. \ref{adapter}. The modeling of the adapter (denoted by C) was not straightforward and hence a transformer---instead of a direct transition---was designed. The transformer (denoted by B) plays the role of a tapered impedance matching network that gradually transform the high impedance of the waveguide input to the standard $50\Omega$ SMA input (denoted by A) over the entire X-band. It is worth noting that while the SMA connector (depicted in yellow) has been modeled in CST for simulation purposes, a commercial solution for the same, available as PE4099 from Pasternack \cite{Pasternack} would be deployed in practice. This is the only commercial element used while designing the entire power distribution network.

\par In addition to power division, the other prime function of the proposed network is to provide phase control to each of the waveguides. This control is facilitated by the design of coupling slots in the main waveguide that connects to the branch waveguides. It is evident from \cite{Rengarajan} that a slot oriented at an angle of $45^\circ$ between two coupled waveguides produces maximum coupling of energy from the feed to the branch. However, this result assumes that the waveguides are terminated to an open circuit and there is an $1:1$ coupling, i.e. the coupling occurs between the main feed and a single waveguide. These assumptions do not hold true here, as the waveguides are terminated to a short and there are eight waveguide branches that need to be coupled to the main waveguide feed. Thus, each slot is oriented at a specific angle with respect to the plane of orientation of the main feed waveguide. At this point optimization algorithms are used to target the following idealized design goals:

\begin{itemize}
    \item $S_{1,1}$ $\rightarrow$ -$\infty$ (Perfectly Matched Input)
    \item $[S_{i,1}]$ = -12.04 dB $\forall$ $i\in[2,17]$ (Equal Power Split at Output Ports)
\end{itemize}

\noindent In practice the above stated goals are actualized as input return loss $(S_{11})<-20\mbox{dB}$ and insertion loss at the ports $(S_{i,1}\forall i\in[2,17])$ around $-12\mbox{dB}$ over the entire band of operation. It is important to know that in this case, the output port matching, $[S_{j,j}\forall j\in[2,17]$ is not relevant because of the shorted termination of the waveguides. The main feed can thus be intuitively considered as a feed-line distributing equal-power to eight waveguide cavities, the upper conducting surfaces of which consist of metasurface PCB antennas that radiate the power. The range of values of the optimization variables for the final structure had to be kept within practical tolerances and machining limits. The slot dimensions (length and breadth) along with their angular orientation were chosen as optimization variables. It is worth noting that the slot angles tend to increase towards the shorted end of the main feed waveguide. This is because the power is decaying along the main feed waveguide due to gradual coupling into the successive waveguides. So the incident power at each slot is less than the previous slot and a higher coupling strength is required to couple equal power. The length of the main feed waveguide is mostly dependent on the width of the branch waveguides. The extra terminating length at the end has been obtained from optimization.
The optimized values of the design parameters are presented in Table-\ref{Design Parameters} and the physical dimensions of the structure is illustrated in Fig. \ref{CST_structure}.

\begin{table}[!ht]
\centering
\caption{Design parameters of the proposed structure at $f_0=9.75\mbox{GHz}$} \vspace{0.05in}
\label{Design Parameters}
\resizebox{6.5cm}{!}{%
\begin{tabular}{|l|l|}
\hline
\textbf{Parameters}            & \textbf{Values} \\ \hline
                               &                 \\ \hline
Waveguide Broadside $(a)$            & 2.286 cm        \\ \hline
Waveguide Shortside $(b)$           & 0.3 cm           \\ \hline
Feed Waveguide Length $(L_{feed})$         & 23.28 cm        \\ \hline
Branch Waveguide Length $(L_{Branch})$       & 10 cm           \\ \hline
$\theta_1$ [O]                    & 12.48$^\circ$   \\ \hline
$\theta_2$ [O]                    & 12.55$^\circ$     \\ \hline
$\theta_3$ [O]                    & 14.08$^\circ$     \\ \hline
$\theta_4$ [O]                    & 14.26$^\circ$     \\ \hline
$\theta_5$ [O]                    & 18.43$^\circ$     \\ \hline
$\theta_6$ [O]                    & 21.48$^\circ$     \\ \hline
$\theta_7$ [O]                    & 28.00$^\circ$     \\ \hline
$\theta_8$ [O]                    & 52.00$^\circ$     \\ \hline
Slot Length $(l_s)$ [O]                   & 2.25 cm         \\ \hline
Slot Width $(w_s)$  [O]                  & 0.36 cm         \\ \hline
Gap Between Waveguide Branches $(s_{wg})$ & 0.2 cm          \\ \hline
Metal Thickness (t)               & 0.1 cm          \\ \hline
\end{tabular}%
}\\
$[O]$ \textit{denotes the parameter chosen as optimization variable}
\end{table}

\begin{figure}[!ht]
\centering{\includegraphics[width=70mm]{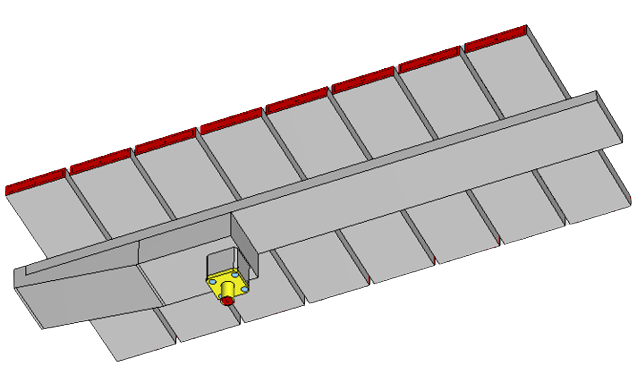}} \vspace{-0.1in}
\caption{3-D view of the composite power distribution network as used for simulation in CST design studio\vspace{-0.1in}
\label{3D_view}}
\end{figure}

\section{Results and Discussion}

\begin{figure}[!ht]
\centering{\includegraphics[width=70mm]{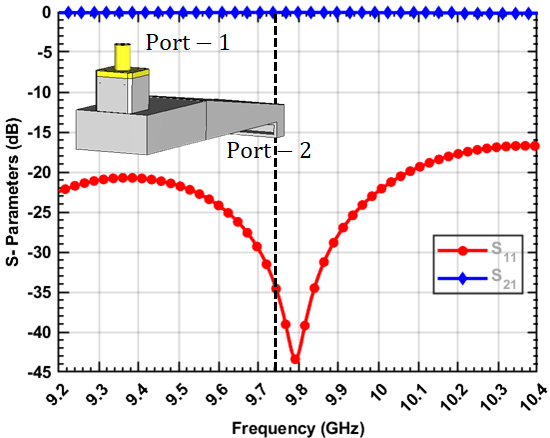}} \vspace{-0.1in}
\caption{\textit{S-parameter response of the SMA-custom waveguide adapter} \vspace{-0.1in}
\label{SMA_adapter}}
\end{figure}

The proposed structure, depicted in Fig. \ref{3D_view}, was implemented and optimized in CST Design Studio. It is worth mentioning that ports 2 through 17 correspond to output ports at either end of each waveguide branch; however, in this work, only the transition has been modeled and the effect of shorted ends has been neglected as an approximation in the simulation. The structure being modular, an S-parameter analysis of the designed SMA-to-custom waveguide adaptor was performed and is presented in Fig. \ref{SMA_adapter}. It can be observed that for all practical purposes, the adaptor exhibits a very good return loss and insertion loss response, indicating excellent matching and negligible loss over the entire band of operation. This confirms the fact that the adaptor, when cascaded with the power distribution network, does not contribute to inherent losses.\vspace{-0.05in}

\par
\begin{figure}[!ht]
\centering{\includegraphics[width=75mm]{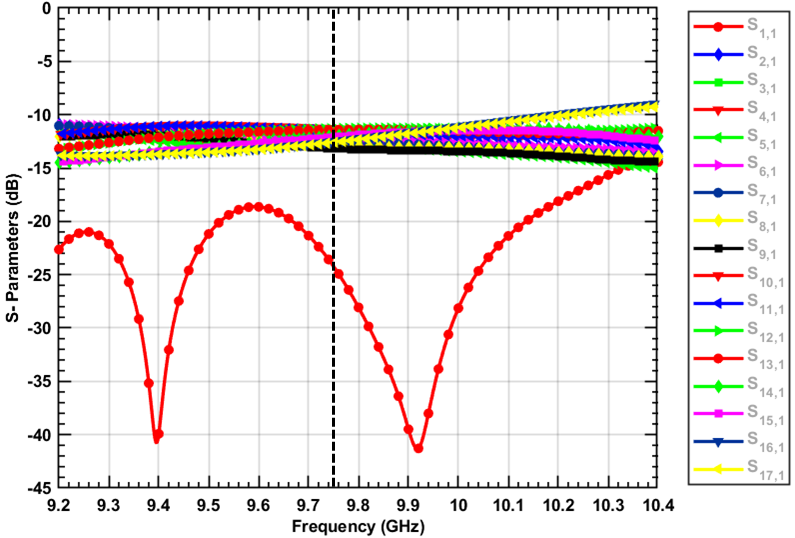}} \vspace{-0.1in}
\caption{S-parameter analysis of the proposed structure indicating equal power split and good match at the desired frequency range
\label{S_results}}
\end{figure}

 The S-Parameter results for the composite structure are shown in Fig. \ref{S_results}. Considering $-15\mbox{dB}$ to be the threshold for acceptable return loss, it can be considered that the proposed power distribution network exhibits a good match over the desired band of operation. The insertion loss is mostly constant with a value of $-12.04\mbox{dB}$ around $9.75\mbox{GHz}$ with a variation of $\pm10\%$. Table-\ref{tab:S_Param} presents the exact values of the S-parameters and phases at the design frequency of $9.75\mbox{GHz}$.

\begin{table}[!ht]
\centering
\caption{S-parameters of the proposed power distribution network}
\label{tab:S_Param}
\begin{tabular}{|l|l|l|}
\hline
\textbf{Parameter}  & \textbf{Value (dB)$^*$} & \textbf{Phase ($^\circ$)$^*$} \\ \hline
           &            &                   \\ \hline
$S_{1,1}$  & -24.23     & N/A               \\ \hline
$S_{2,1}$  & -11.89     & 46.19             \\ \hline
$S_{3,1}$  & -12.22     & -131.43           \\ \hline
$S_{4,1}$  & -12.53     & -176.93           \\ \hline
$S_{5,1}$  & -12.83     & 5.52              \\ \hline
$S_{6,1}$  & -12.45     & -44.36            \\ \hline
$S_{7,1}$  & -12.73     & 138.01            \\ \hline
$S_{8,1}$  & -12.88     & 88.53             \\ \hline
$S_{9,1}$  & -13.16     & -89.53            \\ \hline
$S_{10,1}$ & -11.29     & -135.40           \\ \hline
$S_{11,1}$ & -11.49     & 46.06             \\ \hline
$S_{12,1}$ & -11.30     & 6.74              \\ \hline
$S_{13,1}$ & -11.41     & -171.68           \\ \hline
$S_{14,1}$ & -12.00     & 150.37            \\ \hline
$S_{15,1}$ & -12.07     & -27.74            \\ \hline
$S_{16,1}$ & -12.51     & -79.11            \\ \hline
$S_{17,1}$ & -12.60     & 101.81            \\ \hline
\end{tabular}\\
$^*$ Mentioned at the design frequency of $9.75\mbox{GHz}$
\end{table}

\par
\begin{figure}[!ht]
\centering{\includegraphics[width=75mm]{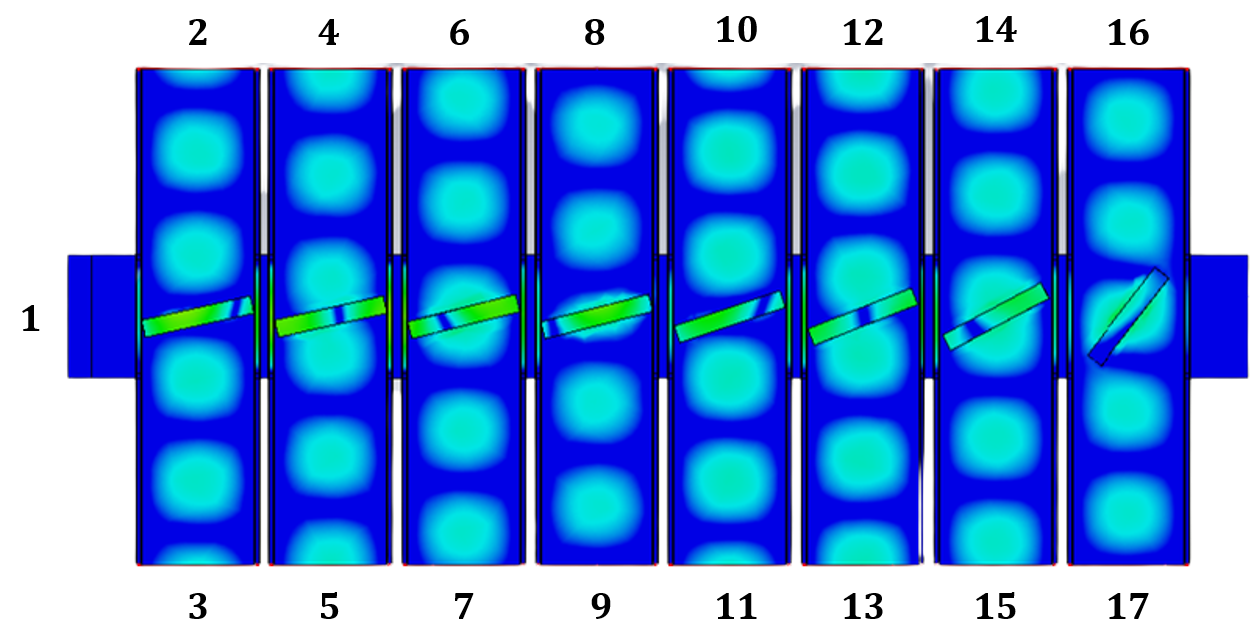}} \vspace{-0.1in}
\caption{Top-view of the E-field phase distribution as power couples through the slots into the waveguides \vspace{-0.4in}
\label{fields}}
\end{figure}

\begin{figure}[!ht]
\centering{\includegraphics[width=75mm]{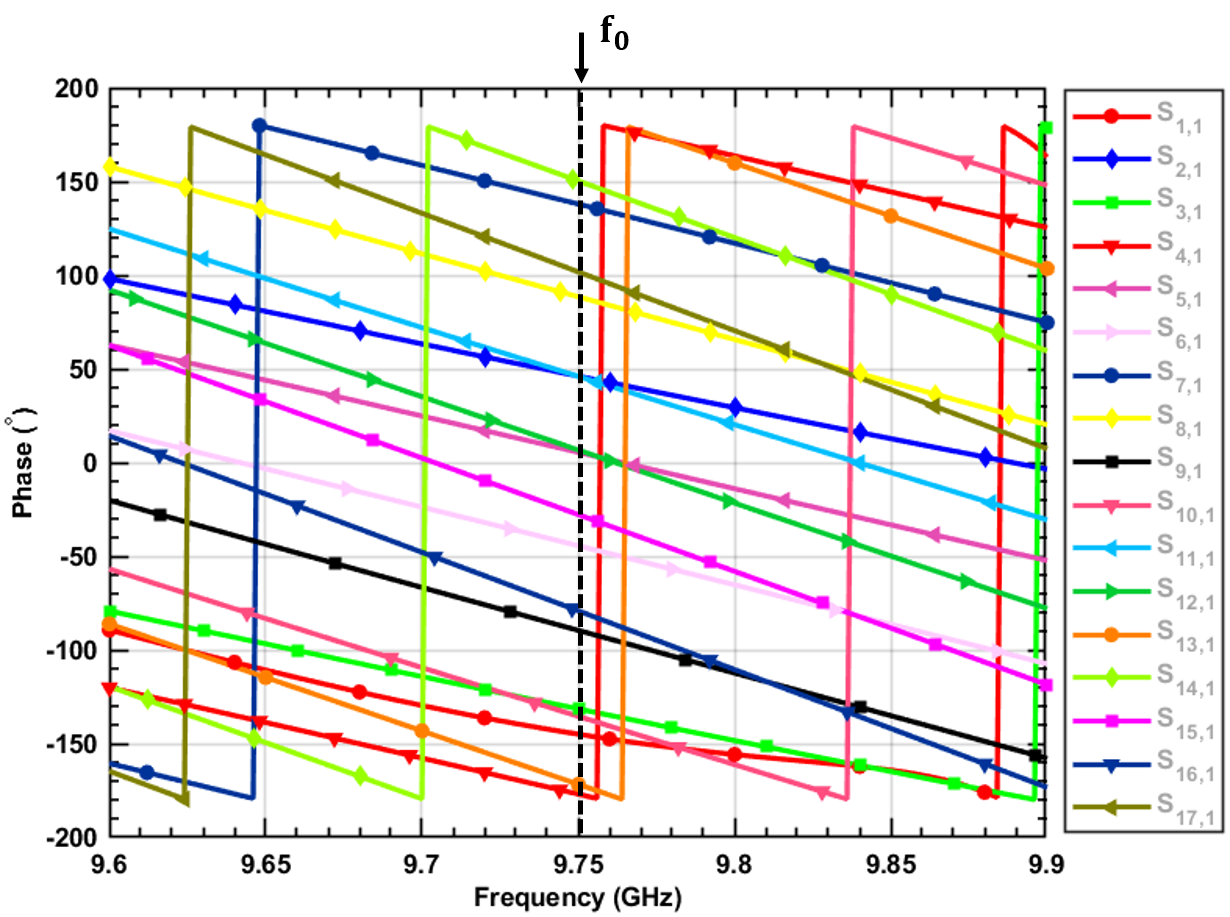}} \vspace{-0.1in}
\caption{Phase distribution of the proposed structure denoting phase diversity \vspace{-0.1in}
\label{phase_results}}
\end{figure}

 The simulated electric field distribution originating from the slots of each waveguide is presented in Fig. \ref{fields}. It is observed that there is a linear phase delay between the wavefronts due to the rotated orientation of the slots. This confirms the differential delay in phase fronts thereby indicating phase diversity in each of the output waveguides. The phase response of the composite power distribution network is depicted in Fig. \ref{phase_results}, which highlights the unique phase transitions for each output ports. It is important to note that the same concept of phase distribution can be implemented using dedicated beamforming networks like a conventional Butler Matrix \cite{Chang} or Rotman Lens \cite{Vashist}; however, such solutions are difficult to implement in X-band, especially when the application requires delivery of hundreds of Watts of power output. Such power levels might preclude the use of PCB laminates to design phase diversification networks, thereby rendering the aforementioned units non-functional. On the other hand, the slot-coupling method is simple to implement on waveguide-based designs, ensuring robust performance for high-power systems and provide phase diversity by varying the angular orientation of the slots. \vspace{-0.2in}

Modern antenna systems of large aperture are often formed from tiling smaller sub-apertures through an efficient power distribution network. Existing commercial power dividers are limited in the number of output ports and multi-way X-band waveguide dividers are extremely costly \cite{Dividers}. As the number of output ports increases, power dividers are often manufactured on a customization basis, implying additional resources and lead-times. An alternative to this involves deploying separate power divider blocks cascaded together in different stages to form a composite power distribution system. Cascading, however, increases the bulk as it requires multiple modules only for a single tile. For multi-tile large aperture systems, this is a limitation as the overall antenna superstructure would require an exceedingly large number of power divider sub-modules in totality. In addition, implementing phase diversity requires separate phase shifters for each branch, thereby further increasing the order of complexity, bulk and connector/transition losses. The presented design avoids these complications. The proposed power distribution network provides a common solution for slot-coupled power to SIW PCB antennas in addition to providing phase diversity. This structure can thus be utilized as a building block for multi-tile antenna superstructures.\vspace{-0.08in}

\section{Conclusion}

A novel simulation-based design methodology for a power distribution network was proposed. The presented design ensures equal power split to eight rows of waveguide branches along with an in-built phase diversification mechanism. The composite feed structure provides an excellent platform for housing metasurface based holographic antennas and aids grating lobe minimization of the same. The design in robust and compact and can be utilized in designing large aperture array antennas for a wide range of applications.\vspace{-0.2in}

\vskip3pt
\ack{This material is based upon work supported by the Defense Advanced Research Projects Agency (DARPA) and Naval Information Warfare Center Pacific (NIWC Pacific) under Contract No. N66001-21-C-4016
}

\vskip5pt

\vskip3pt

\noindent E-mail: todeepayan@gmail.com 

\end{document}